\documentclass[twocolumn,secnumarabic,amssymb, nobibnotes, aps, prb]{revtex4}
\usepackage{graphicx}

\begin{document}

\title{\bf Grain Boundary Diffusion in Copper under Tensile Stress}

\author{Kevin M. Crosby}%
\affiliation{Departments of Physics and Computer Science, Carthage
College, Kenosha, WI 53158}
\date{June 30, 2003}
\pacs{66.30.Fq, 61.72.Mm, 61.72.Ji, 66.30.Pa, 68.35.Fx}
\begin{abstract}
Stress enhanced self-diffusion of Copper on the $\Sigma$3 twin
grain boundary was examined with molecular dynamics simulations.
The presence of uniaxial tensile stress results in a significant
reduction in activation energy for grain-boundary self-diffusion
of magnitude 5 eV per unit strain.  Using a theoretical model of
point defect formation and diffusion, the functional dependence of
the effective activation energy $Q$ on uniaxial tensile strain
$\epsilon$ is shown to be described by
$Q(\epsilon)=Q_0-E_0V^*\epsilon$ where $E_0$ is the
zero-temperature Young's modulus and $V^*$ is an effective
activation volume. The simulation data agree well with this model
and comparison between data and model suggests that
$V^*=0.6\Omega$ where $\Omega$ is the atomic volume.
$V^*/\Omega=0.6$ is consistent with a vacancy-dominated diffusion
mechanism.

\end{abstract}

\maketitle

\def\bd{\begin{displaymath}}
\def\be{\begin{equation}}
\def\ed{\end{displaymath}}
\def\ee{\end{equation}}

\section {Introduction}

Studies of failure mechanisms in nano-structured metallic
interconnects show that diffusion near grain boundaries (GBs)
plays a critical role in initiation of destructive mass-transport.
\cite{turnbull,achter,barnes} For example, electromigration
failure of poly-crystalline aluminum and copper interconnects is
initiated predominantly at grain boundaries and triple points. In
accord with these phenomenological observations are measured and
calculated activation energies for self-diffusion that are
approximately one-third to one-half the corresponding bulk
activation energies.\cite{nomura-jmr-7}  Typically metallic
interconnects are under large tensile strains due to both the
deposition process and the interaction with an underlying
substrate or overlying passivation layer.  Metallic interconnects
in very- and ultra-large-scale-integrated circuit (VLSI/ULSI)
applications can experience tensile stresses in excess of several
hundred MPa resulting in strains on the order of several
hundredths of a percent. While it is well known that external
tensile strain enhances diffusivity in bulk materials, there have
been relatively few experimental or computational investigations
of diffusion in strained metals despite the technological
significance of the problem.

Due to its use in nano-scale interconnect materials, the
structural properties and self-diffusion mechanisms in copper near
GBs have received considerable
attention.\cite{swygenhoven,nomura-jmr-6,nomura-jmr-10,sommer,ma,sorensen}
Several experiments on the microstructure of copper thin films
suggest that the $\Sigma$3 coherent twin grain boundary accounts
for roughly 42\% of all coincident-site-lattice (CSL) boundaries
in as-deposited copper thin films.\cite{hkl}  CSL boundaries
together represent the majority of GB geometries in
poly-crystalline copper. Further, calculations of GB surface
energies show that the $\Sigma 3$ boundary has a markedly lower
energy than all but the $\Sigma 7$ boundary.\cite{rollett} While,
compared to other CSL boundaries, defect mobility is relatively
low in the $\Sigma$3 GB, the large number of such boundaries
present in copper thin films motivates the study of diffusion near
these structures.

Bulk self-diffusion in metals may be considerably simpler than
diffusion near GBs.  Sandberg {\it et al.} have recently
demonstrated that monovacancies are the only active contributors
to bulk diffusion in aluminum at all temperatures up to the
melting point.\cite{sandberg}  The situation near GBs is far from
resolved, however, and is certainly more complicated than that in
the bulk. Sorensen {\it et al.} established that symmetric tilt
GBs support both vacancy and interstitial-mediated diffusion at
low temperatures, while at elevated temperatures, more complex
defect interactions such as Frenkel pairs and even ring mechanisms
contribute to GB diffusion.\cite{sorensen}  Recently, Suzuki and
Mishin demonstrated that vacancy and interstitial mechanisms can
have comparable formation energies in symmetric tilt boundaries in
copper.\cite{suzuki} There is also ample experimental evidence of
complex, cooperative motion of large numbers of atoms in both
twist and tilt GBs in thin Au films at high
temperature.\cite{merkle}

Nomura and Adams have extensively studied diffusion mechanisms in
and near both symmetric tilt and pure twist GBs in
copper.\cite{nomura-jmr-7}  Assuming only a vacancy mechanism for
diffusion in twist boundaries, Nomura and Adams found that
diffusion at low temperatures is dominated by migration along the
screw dislocations comprising high angle twist boundaries, while
high temperature diffusion occurred primarily through the bulk.
Nomura and others have also established that vacancies are
strongly bound to the GB itself so that vacancy diffusion is
primarily restricted to the plane defined by the GB.

Because of the constraining geometry of the twin boundary studied
in the present work, it is unlikely that interstitials and more
complex defects such as Frenkel pairs contribute significantly to
diffusion on the $\Sigma$3 GB at low
temperatures.\cite{nomura-jmr-6} At elevated temperatures,
however, defect concentrations may be high enough that
interactions between point defects may contribute to diffusion
near the GB.  While appropriate mechanisms have been tentatively
identified for bulk diffusion, mechanisms for diffusion near grain
boundaries have not been clearly established.\cite{farakas}

As a computational problem, diffusion near GBs in copper has been
examined in the context of molecular dynamics (MD), molecular
statics (MS), and, more recently, kinetic Monte Carlo (KMC)
models.\cite{nomura-jmr-7,nomura-jmr-6, swygenhoven,sorensen}
Because MD evolves the atomic lattice directly according to a
specified potential, all migration pathways and diffusion
mechanisms are allowed.  For this reason, MD is useful in
identifying the kinetics of defect formation, differentiating
between diffusion through vacancy and interstitial migration when,
as is the case for GB diffusion, a dominant mechanism has not been
identified.

Typically, MD simulations are limited to time scales on the order
of nano-seconds, often too short to reliably extrapolate diffusion
constants at the relatively low (in terms of bulk melting
temperatures) operating temperatures of copper interconnects in
micro-electronics technologies. As a result, an accurate
estimation of activation energies for self-diffusion is difficult.
Conversely, the MS and KMC methods and other approaches involving
the coupled application of Monte Carlo and molecular statics
approaches requires an {\it a priori} knowledge of defect types,
concentrations, and jump frequencies, but overcomes the short-time
scale limitations of direct MD simulations. Recently, hybrid
approaches have been demonstrated that combine the strengths of
each simulation technique to more reliably estimate diffusion
properties in metals.\cite{sorensen}

In this paper, I report on the use of MD to determine activation
energies as a function of static lattice-strain for self-diffusion
on the $\Sigma$3 coherent twin grain boundary in copper. In Sec.
II, the details of the computational model are developed, and
fundamental elastic properties of the simulated crystal are
calculated and compared with existing data to validate the bulk
mechanical behavior of the computational model.  A theoretical
framework for the diffusion data obtained in the simulations is
presented in Sec. III, while in Sec. IV, the activation energy
data and other relevant results are presented and interpreted
within the theoretical framework of Sec. III. Finally, a
discussion of the results in the context of existing and future
experimental and numerical work is presented in Sec. V.

\section{Computational Model}
Molecular dynamics simulations using potentials derived from the
embedded-atom model (EAM) are well-known to accurately account for
the behaviors of metals, particularly near surfaces and point
defects where other semi-classical potentials fail.  The accuracy
of several of these potentials in accounting for the
thermodynamics and energetics of extended defects such as grain
boundaries has recently been established as well.\cite{liu} The
simulations reported here make use of EAM potentials for copper
published by Johnson.\cite{johnson} The potential has been
modified for a smooth cut-off and to better match bulk mechanical
properties of copper.\cite{oh}  The MD time-step is $0.9$ fs, and
each simulation is carried out for at least 1.2 ns, with lower
temperature simulations run for up to 2.2 ns to obtain reasonable
diffusion statistics.

\begin{figure}
  \includegraphics[width=8cm]{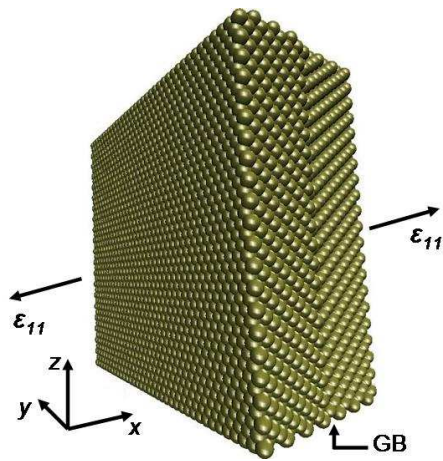}
  \caption{A 12\AA$\times$25\AA$\times$25\AA \space section of the full simulation cell with
  a $\Sigma 3$ GB on the central (111) plane.  A tensile strain is applied along
  the external coordinate direction {\bf x} which is aligned along the
  $\langle 111\rangle$ crystallographic direction.}\label{fig:cell}
\end{figure}

A simulation cell with dimensions 20\AA$\times$40\AA$\times$40\AA,
consisting of roughly 128,000 atoms is sliced in half such that
the slice-plane separates two \{111\} crystal faces of the copper
lattice.  The slice plane defines a grain boundary of area
$1600\mathrm{\AA^2}$. To generate the coherent twin boundary, one
side of the crystal is rotated by $60^\circ$ about the $\langle
111\rangle$ axis. The geometry is illustrated in Fig.
\ref{fig:cell}. Periodic boundary conditions are applied to the
faces of the crystal normal to the GB, while the two surfaces
parallel to the GB are free.  As indicated in Fig. \ref{fig:cell},
the external coordinate direction $\mathbf{x}$ is aligned with the
crystallographic direction $\langle 111\rangle$.

To find the energetically favorable separation of the two
crystallites defining the GB, Monte Carlo relaxation is carried
out to optimize the separation between the two crystallites and a
minimum-energy separation is obtained for each temperature. The
equilibrium lattice constant is established through short MD
calculations which search for a minimum in internal stress as the
lattice constant is varied at a fixed temperature.

For temperatures in the range $0.44T_m <T < 0.74 T_m$, where
$T_m=1356$K is the experimental bulk melting temperature of
copper, the Young's modulus, $E=\sigma_{11}/\epsilon_{11}$ and the
Poisson ratio, $\nu=\epsilon_{22}/\epsilon_{11}$ are evaluated
from the elastic coefficients $C_{11}$ and $C_{12}$. The $C_{ij}$
are obtained from the ensemble averages of fluctuations in the
stress tensor using the method of Ref. 20. Poisson's ratio $\nu$
for tension in the $\mathbf{x}$ direction is relatively
insensitive to temperature and takes the value $\nu=0.31\pm 0.02$
for all simulations reported here.

\begin{figure}
  \includegraphics[width=8cm]{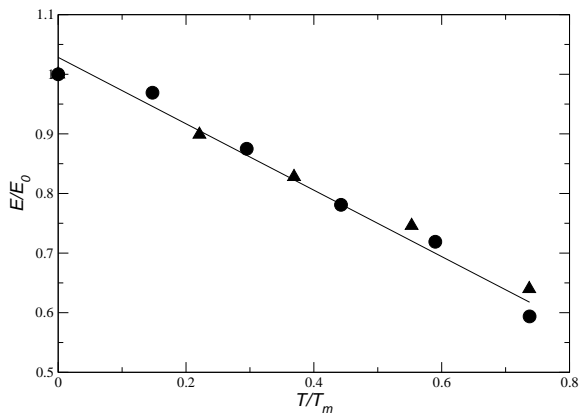}\\
  \caption{The temperature dependence of the Young's modulus
  $E\equiv\sigma_{11}/\epsilon_{11}$ for uniaxial tension in
  units of the zero temperature modulus $E_0=116$ GPa (circles).  The
  solid line is a linear regression fit with slope
  $-0.55$. The triangles are data obtained from Ref. \cite{cagin} with
  $E_0=127$ GPa.  Both data sets are well described by
  $E(T)=E_0(1-\gamma T/T_m)$ with $\gamma=0.55$ and $T_m=1356K$.}
  \label{fig:E}
\end{figure}

The calculated temperature dependence of the Young's modulus
$E(T)$ is shown in Fig. \ref{fig:E} and obeys the phenomenological
relation \be E(T)=E_0(1-\gamma T/T_m)\label{EofT}\ee for
temperatures $T\le 0.74 T_m$ where $E_0=116$ GPa is the
zero-temperature modulus, and $\gamma\approx 0.55$. \c{C}a\u{g}in
{\it et al.} calculate the elastic constant $C_{11}$ as a function
of temperature for copper.\cite{cagin} Their data is in excellent
agreement with Eq. \ref{EofT} when one applies the relation
$C_{11}=(1-\nu)E/(1+\nu)(1-2\nu)$ to obtain the Young's modulus
for uniaxial tension along the $x$ direction from the data in Ref.
\cite{cagin}.

To study diffusion under strain, the lattice is equilibrated at
each target temperature using Monte Carlo relaxation.  A strain
rate of $\dot{\epsilon}=10^{-4} \mathrm{ps}^{-1}$ is applied to
the (111) faces of the lattice until a desired average internal
static strain $\epsilon_{11}$ is reached. Diffusivities are
calculated from the mean-square displacement of atoms in and near
the GB according to \be D_{\alpha}={1\over 6 t}\langle
r_{\alpha}^2(t)\rangle\label{Dmsd}\ee where $r_{\alpha}(t)$ is the
$\alpha$ component of the atomic displacement vector on the GB
plane at time $t$, and the average is over all atoms originating
in either of the two atomic planes defining the grain boundary.
The diffusivity $D_{\parallel}$ parallel to the GB is computed at
temperatures ranging from $0.44 T_m$ to $0.74 T_m$ for a range of
static strains from 0 to 0.07. The maximum corresponding stresses
are on the order of a few GPa.  From the diffusivity data,
Arrhenius plots are constructed and the effective activation
energy is computed at each value of the static strain $\epsilon$.

\section{Diffusion under uniaxial tensile strain}

In the simulations reported here, the copper crystal is under
uniaxial tension $\sigma_{11}$ due to a static strain applied at
the (111) faces as shown in Fig. \ref{fig:cell}. The lattice
strain is given by $\epsilon_{11}=\epsilon$,
$\epsilon_{22}=\epsilon_{33}=-\nu\epsilon$, and all other
$\epsilon_{ij}=0$.  Diffusion in the plane of the GB is
characterized by the diffusivity components $D_{12}$ and $D_{13}$
which we can take to be equal: $D_{\parallel}\equiv
D_{12}=D_{13}$.  Transport normal to the GB is described by the
diffusion constant $D_{11}$. In the simulations reported in this
paper, we have adopted a very narrow definition of the grain
boundary region as consisting of just the two atomic planes
adjacent to the interface. For this reason, we are interested in
the diffusivity $D_{\parallel}$ within the grain boundary plane
$y-z$, and we anticipate that $D_{11}$ will reflect bulk
diffusivity that is not of interest here.

Experimentally, the effect of tensile strain $\epsilon$ on
diffusivity $D$ has been characterized by an effective reduction
of the activation energy for diffusion by the amount\cite{aziz-1,
moriya} \be Q'=-kT{\partial \ln
D\over\partial\epsilon}\label{Qprime}.\ee Aziz has provided a
rigorous theoretical foundation for the empirical result of Eq.
\ref{Qprime}.\cite{aziz-2,aziz-3} We make use of the theory in
Ref. \cite{aziz-3} in what follows.

After Aziz, we consider the components of the diffusivity tensor
$\mathbf{D}$ to be proportional to the product of the defect
concentration $C(\mbox{\boldmath$\sigma$})$ and appropriate
component of the mobility tensor $\mathbf
M(\mbox{\boldmath$\sigma$})$ of point defects in the lattice.
Here, \boldmath$\sigma\mbox{ }$\unboldmath\space is the spatially
uniform stress tensor with components $\sigma_{ij}$. The
equilibrium concentration of defects is related to the stress-free
concentration $C(0)$ by \be {C(\mbox{\boldmath$\sigma$})\over
C(0)}={\exp\left(\mbox{\boldmath$\sigma$}\cdot\mathbf{V^f}\right)\over
kT}.\label{CofSigma}\ee The formation strain tensor ${\mathbf
V^f}$ depends on the defect mechanism, as well as on the location
of the defect source and represents the volume changes due to the
formation of a defect. Experimentally, only certain components of
$\mathbf{V^f}$ are accessible.  In the case of hydrostatic
stresses, the formation volume $V^f\equiv\mathrm{Tr}\mathbf{V^f}$
is measured.

Consider point defects forming on or near the GB plane.  A vacancy
(interstitial) defect initially increases (decreases) the sample
volume by one atomic volume unit $\Omega$ on the (111) face which
is normal to the external coordinate direction $\mathbf{x}$.
Subsequent to a vacancy (interstitial) formation, the lattice
relaxes around the defect, contracting (expanding) around the
vacancy (interstitial).  The resulting change in volume is
characterized by the relaxation volume $V^r$ which is negative
(positive) for vacancy (interstitial) defects. The defect
formation tensor is then given by \be \mathbf{V^f}=\pm\Omega\left(
\begin{array}{ccc}
1 & 0 & 0 \\
0 & 0 & 0 \\
0 & 0 & 0 \end{array} \right)+{V^r\over 3}\left(
\begin{array}{ccc}
1 & 0 & 0 \\
0 & 1 & 0 \\
0 & 0 & 1 \end{array} \right)\label{Vf}\ee where $+$ ($-$) sign is
for vacancy (interstitial) formation.

The directional dependence of the diffusivity is contained in the
mobility tensor, $\mathbf M$.  The defect mobility is also stress
dependent, and its contribution to the self-diffusivity is through
the migration strain tensor $\mathbf{V^m}$.   Diffusion in the
plane of the grain boundary depends on the mobility
$\mathbf{M}_\parallel\equiv\mathbf{M_{12}}\approx\mathbf{M_{13}}$
according to \be
{\mathbf{M_\parallel}(\mbox{\boldmath$\sigma$})\over\mathbf{M_\parallel}(0)}=
{\exp\left(\mbox{\boldmath$\sigma$}\cdot\mathbf{V_{\parallel}^m}\right)\over
kT}\label{MofSigma}\ee where $\mathbf{V_{\parallel}^m}$ is the
migration strain tensor in the plane of the GB and the migration
volume $V^m=\mathrm{Tr}\mathbf{V^m}$ represents the change in
volume of the lattice after the defect has reached the saddle
point in its migration path.

From Equations \ref{CofSigma} and \ref{MofSigma}, the stress
dependence of $D_{\parallel}$ is given by
\be\mathrm{ln}{D_{\parallel}(\mbox{\boldmath$\sigma$})\over
D_{\parallel}(0)}={\mbox{\boldmath$\sigma$}\cdot(\mathbf{V^f}+\mathbf{V_{\parallel}^m})\over
kT} \label{DofSigma}.\ee For the simple tension condition used in
our simulations, the stress tensor has a single non-zero
component, $\sigma_{11}=E\epsilon$ where $\epsilon$ is the
magnitude of the strain along the $x$ direction:
$\epsilon_{11}=\epsilon>0$. Under uniaxial tension in the
$\mathbf{x}$ direction, Eq. \ref{DofSigma} becomes
\be\mathrm{ln}{D_{\parallel}(\epsilon)\over D_{\parallel}(0)}={E
(\pm\Omega+V^r/3+V_{xx}^m)\epsilon\over kT}, \label{DofSigma2}\ee
where $V_{xx}^m$ is the dimension change in the $\mathbf{x}$
direction upon defect migration to its saddle point.

In an unstrained lattice, the usual Arrhenius form for the
diffusivity, \be \mathrm{ln}{D\over D_0}= -{Q_0\over
kT}\label{D}\ee holds, where $D_0$ is a material constant, and
$Q_0$ is the activation energy for self-diffusion in the
unstrained lattice. $Q_0$ is the sum of a defect formation energy
$Q_0^f$ and a migration energy, $Q_0^m$. It is clear from Eqns.
\ref{DofSigma2} and \ref{D} that the effective activation energy
for the case of simple tension considered here is \be
Q(\epsilon)=Q_0-EV^*\epsilon\label{Q},\ee where \be
V^*\equiv\pm\Omega+V^r/3+V_{xx}^m.\label{V*}\ee The effect of
tensile strain is to reduce the effective activation energy for
self-diffusion by the amount $EV^*$ per unit strain.

High-temperature elastic softening of the copper crystal does not
change the fundamental Arrhenius analysis of diffusion
coefficients. Combining Eqns. \ref{EofT}, \ref{DofSigma}, and
\ref{V*}, we find the Diffusion coefficient is still of Arrhenius
form: \be \mathrm{ln}{D_{\parallel}(\epsilon)\over
\tilde{D}_{\parallel\mbox{}0}(\epsilon)}=-{Q(\epsilon)\over
kT}\label{QofEpsilon}\ee where $\tilde{D}_{\parallel\mbox{
}_0}(\epsilon)=D_{\parallel\mbox{}0}e^{-\gamma E_0
V^*\epsilon/kT_m}$, and $D_{\parallel\mbox{}0}$ is a material
constant. The strain-dependent activation energy $Q(\epsilon)$ in
Eq. \ref{QofEpsilon} is given by Eq. \ref{Q} with $E$ replaced by
$E_0$. The functional form of the effective activation energy is
unchanged by elastic softening, and the usual Arrhenius analysis
is still appropriate for the determination of diffusivities and
activation energies.

\section{Simulation Results}
\begin{figure}
  \includegraphics[width=8cm]{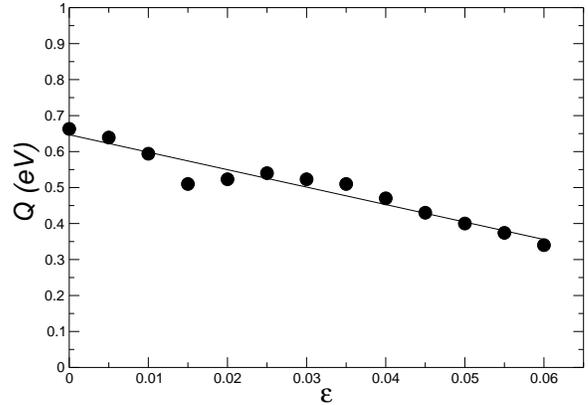}\\
  \caption{The effective activation energy is plotted against the strain
  $\epsilon_{11}=\epsilon$.  The solid line is a linear regression fit
  with slope 4.9 eV.}\label{fig:Qeff}
\end{figure}

The essential result of the MD simulations is the functional
dependence of effective activation energy on lattice strain for
diffusion on the GB plane. The function $Q(\epsilon)$ is obtained
from the Arrhenius plots of Diffusivity $D_{\parallel}$ vs.
inverse temperature.  This dependence is illustrated in Fig.
\ref{fig:Qeff}. The solid line in Fig. \ref{fig:Qeff} is a linear
regression fit to the data of the form
$Q(\epsilon)=0.65\mathrm{eV}-4.9\mathrm{eV}\epsilon$. While
experimental data with conditions similar to those simulated here
are not known to us, there are some general observations and
comparisons that can be made with respect to the functional
dependence of activation energy on strain.  As expected, the
zero-strain activation energy $Q(0)=0.65$ eV is higher than that
for most tilt boundaries in copper, but significantly lower than
the bulk activation energy of 1.3 eV for self-diffusion in copper.

To place our data in the context of the analysis in Sec. III, let
us compare the slope of the $Q(\epsilon)$ data in Fig.
\ref{fig:Qeff} with the predicted slope $E_0V^*$ (Eq. \ref{Q}).
With $E_0=116$ GPa, we find $V^*=0.6\Omega$, where
$\Omega=1.2\times 10^{-29}m^3$ is the atomic volume for
fcc-copper.  While there is no experimental data with which to
compare the value of $V^*$, $V^*/\Omega=0.6$ is consistent with
vacancy-dominated diffusion and inconsistent with an
interstitial-type mechanism.  This conclusion is based on the
assumption that the migration volume contribution $V^*_{xx}$ is
negligible in comparison to the other volume terms in Eq.
\ref{V*}.  Assuming a vacancy mechanism in Eq. \ref{V*}, we find
that the relaxation volume satisfies $V^r/\Omega=-1.2$, while the
assumption of a pure interstitial mechanism results in
$V^r/\Omega=+4.8$.  The latter value is implausibly large, and we
conclude that simple vacancies represent the dominant point
defects in the GB.

\section{Discussion}

To estimate the technological significance of strain in
interconnect device applications, consider that tensile stresses
in copper interconnects are typically on the order of several
hundred MPa, and the resulting strains depend on the
crystallographic direction of the applied stress, but are on the
order of $10^{-3}$. A reduction in effective activation energy of
$\approx 5$ eV per unit strain suggests that the strain effect is
of order $10^{-2}$ eV, effectively doubling diffusivity at room
temperature.  While experimental diffusivity measurements can be
uncertain to within orders of magnitude, static intrinsic strains
typical of modern interconnect materials may significantly enhance
diffusion near GBs in these materials.

Any attempt to calculate activation energies from diffusion
constants obtained with MD simulations is risky, in that the
simulation time-scale may be shorter than the time taken for the
defect concentration to reach equilibrium. However, the
zero-strain activation energy $Q_0=0.65$ eV reported in this paper
is within the range of values obtained by others for diffusion on
a variety of GBs using molecular statics and assumptions about the
dominant defect mechanism.\cite{nomura-jmr-7}  The evidence for a
dominant vacancy-type mechanism is also consistent with several
previous studies of defect formation on coherent twin GBs.

Future computational work on grain boundary diffusion in copper
will focus on measurements of $V^*$ for other technologically
important GBs such as the $\Sigma 7$ GB family.  Like the $\Sigma
3$ twin, $\Sigma 7$ GBs are found in relatively high numbers in
as-deposited copper thin films, but typically have significantly
higher defect mobility than the $\Sigma 3$ GB.


\begin{thebibliography}{99}

\bibitem{achter} M. R. Achter and R. Smoluchowski, J. Appl. Phys.
{\bf 22}, 1260 (1951).
\bibitem{barnes} R. S. Barnes, Nature {\bf 166}, 1032 (1950).
\bibitem{turnbull} D. Turnbull, Phys. Rev. {\bf 76}, 417A (1949).
\bibitem{nomura-jmr-7}  M. Nomura and J. Adams, J. Mater. Res.
{\bf 7}, 3202 (1992).
\bibitem{nomura-jmr-6} M. Nomura, S. Lee, and J. Adams, J. Mater.
Res. {\bf 6}, 1 (1991).
\bibitem{sorensen} M. Sorensen, Y. Mishin, and A. Voter, Phys.
Rev. B {\bf 62}, 3658 (2000).
\bibitem{swygenhoven} H. Van Swygenhoven, D. Farakas, and A. Caro,
Phys. Rev. B {\bf 62}, 831 (2000).
\bibitem{nomura-jmr-10} M. Nomura and J. Adams, J. Mater. Res., {\bf 10}, 2916 (1995).
\bibitem{ma} Q. Ma and R. W. Balluffi, Acta Metall. Mater. {\bf 41}, 133 (1993).
\bibitem{sommer} J. Sommer, Chr. Herzig, T. Muschik, and W. Gust, Acta Metall.
Mater. {\bf 43}, 1099 (1995).
\bibitem{hkl} {\it Grain size, grain boundary and quantitative
texture analysis of a Cu thin film} (2003), from HKL Technologies
Application Notes:
http://www.hkltechnology.com//applic\_notes//app7.pdf.
\bibitem{rollett} A. Rollett, Mat. Sci. Forum:
Proceedings of the 8th International Conference on Aluminum and
its Alloys {\bf 396}, 593 (2002).
\bibitem{sandberg} N. Sandberg, B. Magyari-K\"{o}pe, and T.
Mattsson, Phys. Rev. Lett. {\bf 89}, 65901 (2002).
\bibitem{suzuki} A. Suzuki and Y. Mishin, Interface Science {\bf 11}, 131
(2003).
\bibitem{merkle} K. L. Merkle, L. J. Thompson, and F. Phillipp,
Phys. Rev. Lett. {\bf 88}, 225501 (2002).
\bibitem{farakas} D. Farakas, J. Phys.: Condens. Matter {\bf 12}, R497 (2000).
\bibitem{liu} C. Liu and S. J. Plimpton, J. Mater. Res. {\bf 10}, 1589
(1995).
\bibitem{johnson} R. A. Johnson, Phys. Rev. B {\bf 37}, 3924
(1988).
\bibitem{oh} D. J. Oh and R. A. Johnson, J. Mater. Res. {\bf
3}, 471 (1996).
\bibitem{ray} J. Ray and A. Rahman, J. Chem. Phys. {\bf 80}, 4243
(1985).
\bibitem{cagin} T. \c{C}a\u{g}in, G. Dereli, M. Uludo\u{g}an, and
M. Tomak, Phys. Rev. B {\bf 59}, 3468 (1999).
\bibitem{aziz-1} M. Aziz, {\it Defect and Diffusion Forum}
{\bf 153-155}, 1 (1998).
\bibitem{moriya} N. Moriya, L.C. Feldman, H. S. Luftman, C. A.
King, J. Bevk, and B. Freer, Phys. Rev. Lett. {\bf 71}, 883
(1993).
\bibitem{aziz-2} M. J. Aziz, P. C. Sabin, and G. Lu, Phys. Rev. B
{\bf 44}, 9812 (1991).
\bibitem{aziz-3} M. J. Aziz, Appl. Phys. Lett. {\bf 70}, 26 (1997).

\end{thebibliography}
\end{document}